\documentclass[preprint,12pt]{elsarticle}

\usepackage{amssymb}
\usepackage{amsmath}

\usepackage{ucs}
\usepackage[utf8]{inputenc}
\usepackage[english]{babel}
\usepackage{fontenc}
\usepackage{graphicx}
\usepackage{float}
\usepackage{color}
\usepackage{amsmath, amsthm, amsfonts}
\usepackage{braket}
\usepackage{amsmath}
\usepackage{amsfonts}
\usepackage{mathabx}
\usepackage{amssymb}
\usepackage[toc,page]{appendix}
\usepackage{xcolor}
\usepackage{graphicx}
\usepackage{natbib}
\usepackage{comment}
\usepackage{xpatch}
\usepackage{filecontents,verbatim}

\usepackage{float}
\usepackage[caption = false]{subfig}

\usepackage[symbol]{footmisc}
\usepackage[colorlinks=true]{hyperref}
\newcommand{\be}{\begin{equation}}
\newcommand{\ee}{\end{equation}}
\newcommand{\ba}{\begin{eqnarray}}
\newcommand{\ea}{\end{eqnarray}}
\newcommand{\ma}{\Delta m^2_{31}}

\newcommand{\sa}{\sin^2\theta_{23}}

\newcommand{\dcp}{\delta_{\textrm{CP}}}

\def\eV{\rm{eV}^2}

\newcommand{\dmf}{\Delta m^2_{41}}

\newcommand{\capdef}{}
\newcommand{\mycaption}[2][\capdef]{\renewcommand{\capdef}{#2}
	\caption[#1]{{\footnotesize #2}}}
\makeatletter
\renewcommand{\fnum@table}{\textbf{\tablename~\thetable}}
\renewcommand{\fnum@figure}{\textbf{\figurename~\thefigure}}
\makeatother

\makeatletter
\def\ps@pprintTitle{%
	\let\@oddhead\@empty
	\let\@evenhead\@empty
	\def\@oddfoot{\centerline{\thepage}}%
	\let\@evenfoot\@oddfoot}
\makeatother

\usepackage{float}

\begin{document}

\begin{frontmatter}



\title{Sensitivity to light sterile neutrino mixing parameters with KM3NeT/ORCA}


% ----- Start automatically generated KM3NeT info
% ----- Start author list

\cortext[cor]{corresponding author}

\author[a]{S.~Aiello}
\fntext[]{KM3NeT Publication Committee: km3net-pc@km3net.de.}
\author[bc,b]{A.~Albert}
\author[c]{M.~Alshamsi}
\author[d]{S. Alves Garre}
\author[e]{Z.~Aly}
\author[f,g]{A. Ambrosone}
\author[h]{F.~Ameli}
\author[i]{M.~Andre}
\author[j]{G.~Androulakis\fnref{fn0}}
\fntext[fn0]{Deceased.}
\author[k]{M.~Anghinolfi}
\author[l]{M.~Anguita}
\author[m]{G.~Anton}
\author[n]{M. Ardid}
\author[n]{S. Ardid}
\author[c]{J.~Aublin}
\author[j]{C.~Bagatelas}
\author[c]{B.~Baret}
\author[o]{S.~Basegmez~du~Pree}
\author[c,p]{M.~Bendahman}
% ...
\fntext[fn1]{Presently at the University of Cincinnati, Ohio, United States.}
\author[u]{H.~Thiersen}
\author[al]{S.~Tingay}
\author[j]{S.~Tsagkli}
\author[j]{V.~Tsourapis}
\author[j]{E.~Tzamariudaki}
\author[j]{D.~Tzanetatos}
\author[c,av]{V.~Van~Elewyck}
\author[ba]{G.~Vasileiadis}
\author[q,r]{F.~Versari}
\author[f,ac]{D.~Vivolo}
\author[c]{G.~de~Wasseige}
\author[bb]{J.~Wilms}
\author[at]{R.~Wojaczy\'nski}
\author[o,aa]{E.~de~Wolf}
\author[w]{T.~Yousfi}
\author[k]{S.~Zavatarelli}
\author[ae,h]{A.~Zegarelli}
\author[t]{D.~Zito}
\author[d]{J.\,D.~Zornoza}
\author[d]{J.~Z{\'u}{\~n}iga}
\author[u]{N.~Zywucka}
% ----- End author list
% ----- Start affiliation list
\address[a]{INFN, Sezione di Catania, Via Santa Sofia 64, Catania, 95123 Italy}
\address[b]{Universit{\'e}~de~Strasbourg,~CNRS,~IPHC~UMR~7178,~F-67000~Strasbourg,~France}
\address[c]{Universit{\'e} de Paris, CNRS, Astroparticule et Cosmologie, F-75013 Paris, France}
\address[d]{IFIC - Instituto de F{\'\i}sica Corpuscular (CSIC - Universitat de Val{\`e}ncia), c/Catedr{\'a}tico Jos{\'e} Beltr{\'a}n, 2, 46980 Paterna, Valencia, Spain}
\address[e]{Aix~Marseille~Univ,~CNRS/IN2P3,~CPPM,~Marseille,~France}
\address[f]{INFN, Sezione di Napoli, Complesso Universitario di Monte S. Angelo, Via Cintia ed. G, Napoli, 80126 Italy}
\address[g]{Universit{\`a} di Napoli ``Federico II'', Dip. Scienze Fisiche ``E. Pancini'', Complesso Universitario di Monte S. Angelo, Via Cintia ed. G, Napoli, 80126 Italy}
\address[h]{INFN, Sezione di Roma, Piazzale Aldo Moro 2, Roma, 00185 Italy}
\address[i]{Universitat Polit{\`e}cnica de Catalunya, Laboratori d'Aplicacions Bioac{\'u}stiques, Centre Tecnol{\`o}gic de Vilanova i la Geltr{\'u}, Avda. Rambla Exposici{\'o}, s/n, Vilanova i la Geltr{\'u}, 08800 Spain}
\address[j]{NCSR Demokritos, Institute of Nuclear and Particle Physics, Ag. Paraskevi Attikis, Athens, 15310 Greece}
\address[k]{INFN, Sezione di Genova, Via Dodecaneso 33, Genova, 16146 Italy}
\address[l]{University of Granada, Dept.~of Computer Architecture and Technology/CITIC, 18071 Granada, Spain}
\address[m]{Friedrich-Alexander-Universit{\"a}t Erlangen-N{\"u}rnberg, Erlangen Centre for Astroparticle Physics, Erwin-Rommel-Stra{\ss}e 1, 91058 Erlangen, Germany}
\address[n]{Universitat Polit{\`e}cnica de Val{\`e}ncia, Instituto de Investigaci{\'o}n para la Gesti{\'o}n Integrada de las Zonas Costeras, C/ Paranimf, 1, Gandia, 46730 Spain}
\address[o]{Nikhef, National Institute for Subatomic Physics, PO Box 41882, Amsterdam, 1009 DB Netherlands}
\address[p]{University Mohammed V in Rabat, Faculty of Sciences, 4 av.~Ibn Battouta, B.P.~1014, R.P.~10000 Rabat, Morocco}
\address[q]{INFN, Sezione di Bologna, v.le C. Berti-Pichat, 6/2, Bologna, 40127 Italy}
\address[r]{Universit{\`a} di Bologna, Dipartimento di Fisica e Astronomia, v.le C. Berti-Pichat, 6/2, Bologna, 40127 Italy}
% ...
\address[t]{INFN, Laboratori Nazionali del Sud, Via S. Sofia 62, Catania, 95123 Italy}
\address[u]{North-West University, Centre for Space Research, Private Bag X6001, Potchefstroom, 2520 South Africa}
% ...
\address[w]{University Mohammed I, Faculty of Sciences, BV Mohammed VI, B.P.~717, R.P.~60000 Oujda, Morocco}
% ...
\address[aa]{University of Amsterdam, Institute of Physics/IHEF, PO Box 94216, Amsterdam, 1090 GE Netherlands}
% ...
\address[ac]{Universit{\`a} degli Studi della Campania "Luigi Vanvitelli", Dipartimento di Matematica e Fisica, viale Lincoln 5, Caserta, 81100 Italy}
\address[ae]{Universit{\`a} La Sapienza, Dipartimento di Fisica, Piazzale Aldo Moro 2, Roma, 00185 Italy}
% ...
\address[al]{International Centre for Radio Astronomy Research, Curtin University, Bentley, WA 6102, Australia}
% ...
\address[at]{National~Centre~for~Nuclear~Research,~02-093~Warsaw,~Poland}
% ...
\address[av]{Institut Universitaire de France, 1 rue Descartes, Paris, 75005 France}
% ...
\address[ba]{Laboratoire Univers et Particules de Montpellier, Place Eug{\`e}ne Bataillon - CC 72, Montpellier C{\'e}dex 05, 34095 France}
\address[bb]{Friedrich-Alexander-Universit{\"a}t Erlangen-N{\"u}rnberg, Remeis Sternwarte, Sternwartstra{\ss}e 7, 96049 Bamberg, Germany}
\address[bc]{Universit{\'e} de Strasbourg, Universit{\'e} de Haute Alsace, GRPHE, 34, Rue du Grillenbreit, Colmar, 68008 France}
% ...
% ----- End affiliation list
% ----- End automatically generated KM3NeT info

\begin{abstract}
KM3NeT/ORCA is a next-generation neutrino telescope optimised for atmospheric neutrino oscillations studies. In this paper, the sensitivity of ORCA to the presence of a light sterile neutrino in a 3+1 model is presented. After three years of data taking, ORCA will be able to probe the active-sterile mixing angles $\theta_{14}$, $\theta_{24}$, $\theta_{34}$ and the effective angle $\theta_{\mu e}$, over a broad range of mass squared difference $\dmf \sim [10^{-5}, 10]$ $\eV$, allowing to test the eV-mass sterile neutrino hypothesis as the origin of short baseline anomalies, as well as probing the hypothesis of a very light sterile neutrino, not yet constrained by cosmology. ORCA will be able to explore a relevant fraction of the parameter space not yet reached by present measurements. 
\end{abstract}

\end{frontmatter}

%%\umbers


\section{Introduction}
\label{sec:introduction}
The study of neutrino oscillations has seen remarkable progress in the last three decades. An increasing number of solar, atmospheric and accelerator neutrino experiments have performed precision measurements of the neutrino oscillation parameters \cite{PDG}. The experimental data is consistent with the three weakly-interacting neutrino picture (here referred to as the standard picture). Nevertheless, a number of questions remain unanswered, in particular what is the Neutrino Mass Ordering (NMO) and whether neutrino oscillations violate the CP symmetry. Upcoming experiments such as KM3NeT/ORCA \cite{loi}, SBN \cite{SBN}, DUNE \cite{DUNE}, JUNO \cite{JUNO}, Hyper-K \cite{Hyper-K}, IceCube/Gen2 \cite{ic_gentwo} and INO \cite{INO} aim to resolve these questions over the next decades. 
\\
At the same time, several short baseline (SBL) neutrino experiments have reported anomalous experimental results which are inconsistent with the standard picture. A comprehensive review can be found in Ref. \cite{sterile_review}. Such results could be explained by assuming the existence of an additional neutrino (hereafter SBL neutrino). However, the Z-width measurement \cite{Zwidth} has demonstrated that only three neutrinos can participate to weak interactions, for which they are referred as active neutrinos. Therefore, the SBL neutrino, not being able to participate to weak interactions, is called sterile. The SBL sterile neutrino should be light ($\dmf \sim 1 \,\eV$) and its presence affects the standard neutrino oscillation probabilities via its mixing with active neutrinos, in the so called 3+1 model. 
\\
Specifically, oscillations in the presence of a single sterile neutrino can be modelled by extending the standard picture to include four neutrino eigenstates. In this case, six new parameters are introduced in the model: one additional mass square difference $\dmf$, three active-sterile mixing angles $\theta_{14}$, $\theta_{24}$ and $\theta_{34}$, and two additional CP-violating phases $\delta_{14}$, $\delta_{24}$.
\\
The neutrino evolution in matter can be described by the following effective Hamiltonian:
\begin{equation}
\label{eq:Ham}
H = UH_0U^\dagger + V,
\end{equation}
\noindent where $H_0 = \mbox{diag}(0, \Delta{m^2_{21}}, \Delta{m^2_{31}}, \Delta{m^2_{41}}) / 2E$, and $V = \sqrt{2}G_F\mbox{diag(}N_e, 0, 0, N_n/2)$, with $G_F$ being the Fermi constant and $N_e$, $N_n$ representing the density of electrons and neutrons in the propagation medium. $U$ is an extended $4 \times 4$ unitary matrix relating flavour and mass eigenstates, which can be parametrised such that:
\begin{equation}
U = R_{34} \tilde{R}_{24} \tilde{R}_{14} R_{23} \tilde{R}_{13} R_{12},
\end{equation}
\noindent where $R_{jk}$ is a rotation matrix in the $j\mbox{-}k$ plane and, similarly, $\tilde{R}_{jk}$ is a generalised unitary rotation matrix with an added complex phase.
\\
 In the 3+1 model, the active-sterile mixing elements are expressed by 
\begin{eqnarray}
U_{e4} &=& \sin \theta_{14} e^{-i \delta_{14}}, \\
U_{\mu4} &= & \cos \theta_{14} \sin \theta_{24} e^{-i \delta_{24}}, \\
U_{\tau4} &=& \cos \theta_{14} \cos \theta_{24} \sin \theta_{34}.
\end{eqnarray}
Several experiments have been searching for the SBL sterile neutrino. To date, results are not fully consistent with the 3+1 model: disappearance experiments results are compatible with the standard neutrino scenario while some appearance experiments, such as LSND \cite{LSND} and MiniBooNE \cite{Miniboone}, observed significant $\nu_e$ or $\bar{\nu}_e$ excesses. The global fit of the experimental data with the 3+1 model results in a poor goodness-of-fit, suggesting the need of additional factors in order to explain all data.
\\
Even stronger bounds on the sterile parametric space come from cosmology \cite{strongboundCosmology}, which indirectly constrains the effective number of relativistic species $N_{\rm eff}$ in our Universe. Theoretically, the three active neutrinos give $N_{\rm eff} \sim 3$ \cite{Dolgov}. If a light sterile neutrino with the mixing parameters determined by SBL oscillations is included in the model, it should have been fully thermalised with the active neutrinos \cite{gariazzo2016light}.
This would require $N_{\rm eff} \sim 4$. Cosmological data measure a value of $N_{\rm eff}$ well-compatible with three neutrino species \cite{review_nu_cosmo}, showing a tension with the SBL anomalies. Such a tension is relaxed when cosmological data are combined with astrophysical measurements of cepheids, supernovae and gravitational lensing. In this case, the obtained value of $N_{\rm eff}$ is compatible with four at 68$\%$ C.L. \cite{review_nu_cosmo, gariazzo2016light}.
\\
More generally, cosmological data alone can be compatible with a sterile neutrino with a mass in the eV range only if its contribution to $N_{\rm eff}$ is very small, or with a somewhat larger $N_{\rm eff}$ only if it comes from a nearly massless sterile particle \cite{Planck, Archidiacono_2016}. 
\\
Therefore, more terrestrial and cosmological observations are necessary to understand the origin of the SBL anomalies. Moreover, new observations able to constrain the not-fully-excluded sterile neutrino region from cosmology, at very low sterile mass splittings ($\dmf \ll 1 \, \eV$) can further contribute to testing the sterile neutrino hypothesis.
\\
In this context, the role of next-generation neutrino detectors, such as KM3NeT, is relevant, given their ability to probe the sterile neutrino hypothesis with atmospheric neutrinos \cite{Razzaque_2011, Razzaque_2012}. KM3NeT is a research infrastructure hosting a network of next generation neutrino telescopes currently under construction in the Mediterranean Sea \cite{loi} and built upon the experience from the ANTARES neutrino telescope \cite{antares_loi}. Once completed, KM3NeT will consist of two detectors: (1) ORCA (Oscillation Research with Cosmics in the Abyss) near Toulon, France, optimised for GeV-scale atmospheric neutrino studies, and (2) ARCA (Astroparticle Research with Cosmics in the Abyss), in Sicily, Italy, optimised for the observation of higher-energy ($E_\nu > 1$ TeV) neutrinos from astrophysical sources. 
\\
By exploiting the natural source of atmospheric neutrinos, passing through the Earth and interacting within the detector volume, KM3NeT will perform neutrino oscillation studies over a broad range of energies (from few GeV up to PeV) and baselines (up to the Earth diameter).
Matter effects, experienced by atmospheric neutrinos during their passage through the Earth, are expected to enhance the effect of the presence of a sterile neutrino. Moreover, the wide L/E range available in KM3NeT increases its potential to investigate the existence of a sterile neutrino in the 3+1 model.
\\
This paper is focused on the ORCA capability to search for a light sterile neutrino. It will be shown that ORCA has a high potential to simultaneously constrain the active-sterile mixing angles $\theta_{14}$, $\theta_{24}$, $\theta_{34}$ and the effective angle $\theta_{\mu e}$, with three years of data taking. Particularly, the ORCA sensitivity to such parameters is competitive with other experiments for sterile neutrino mass at the eV scale, indicated by SBL anomalies, and it is able to provide even stronger constraints for extremely low sterile mass splittings ($\dmf$ down to $10^{-5} \, \eV$).
\\
This paper is organised as follows: Section \ref{sec:km3net} describes the KM3NeT/ORCA neutrino telescope. Section \ref{sec:phenomeno} discusses the 3+1 flavour model and oscillation probabilities. Section \ref{sec:analysismethod} describes the sterile neutrino analysis method, including a brief summary of the ORCA Monte Carlo (MC) simulation flow. Results on the ORCA sensitivity are presented in Section~\ref{sec:sensi}. Finally, the results are summarised and discussed in Section~\ref{sec:summary}.

\section{The KM3NeT/ORCA Detector}
\label{sec:km3net}
KM3NeT/ORCA is a deep water neutrino detector under construction in the Mediterranean Sea. Its location is $42^\circ 48'$ N $06^\circ 02'$ E, about 40 km offshore from Toulon, France, at a depth of about 2450 m. Upon its completion, ORCA will consist of 115 flexible detection units (DUs), 200 m high, each comprising 18 Digital Optical Modules (DOMs). A DOM is a pressure resistant, 17-inch diameter glass sphere containing a total of 31, 3" photomultiplier tubes (PMTs) and their associated electronics. 
\\
The primary goal of ORCA is to determine the neutrino mass ordering and to make neutrino oscillation measurements, such as atmospheric parameters ($\sa$, $\ma$) as well as to search for $\nu_\tau$ appearance \cite{ORCA_NMO_Paper}. Neutrino oscillation studies \cite{loi} have demonstrated the presence of a resonance in neutrino oscillation probabilities for few-GeV ($2-8$ GeV) atmospheric neutrinos passing through the Earth. Such a resonance allows the NMO \cite{loi} measurement. 
\\
The ORCA geometrical configuration is optimised for studies with atmospheric neutrinos in the few GeV range: the horizontal spacing between DUs is $\sim 20$ m, whereas the vertical spacing between DOMs in each DU is $\sim 9$ m, with the first DOM being about 30 m above the seabed. The total instrumented volume is $6.7 \cdot 10^6$ m$^3$ (about 7 Mt of sea water). 
\\
In this energy regime, the events produced by atmospheric neutrinos interacting in water are spatially contained. In particular, two event topologies can be produced: track-like events, characterised by a long muon track, mostly from $\nu_\mu$ charged-current (CC) interactions in water, and shower-like events, characterised by events with no distinguishable tracks, mostly from $\nu_e$-CC and all neutral-current (NC) interactions, but with sizeable contributions from $\nu_\tau$-CC and $\nu_\mu$-CC events with short tracks. A track-like event in water has a length of $\sim 4$ m/GeV, whereas shower-like events have a $\log$(E/GeV) dependence, which corresponds to a size of the order of a few meters.
\\
The ORCA detector is an excellent instrument for the sterile neutrino search due to its dense configuration and to matter effects, whose impact in oscillation probabilities of GeV neutrinos travelling in the Earth is described in the next section.
\\
More details on KM3NeT/ORCA can be found in \cite{loi, ORCA_NMO_Paper}.

\newcommand{\UNP}{U^{4\nu}}
\newcommand{\USM}{U^{3\nu}}

\section{Theoretical Background}
\label{sec:phenomeno}
\begin{figure}
    \centering
    \subfloat{\includegraphics[width=0.49\textwidth]{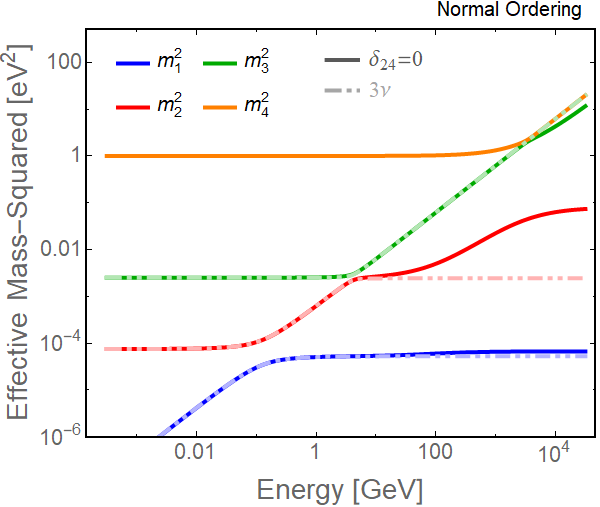}}
    \subfloat{\includegraphics[width=0.49\textwidth]{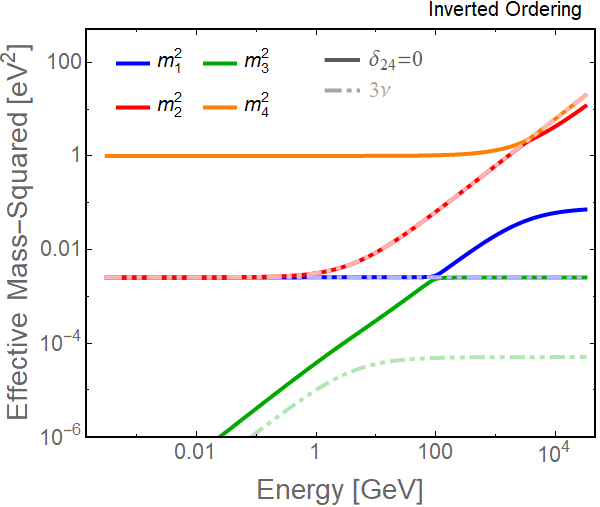}}\\
    \subfloat{\includegraphics[width=0.49\textwidth]{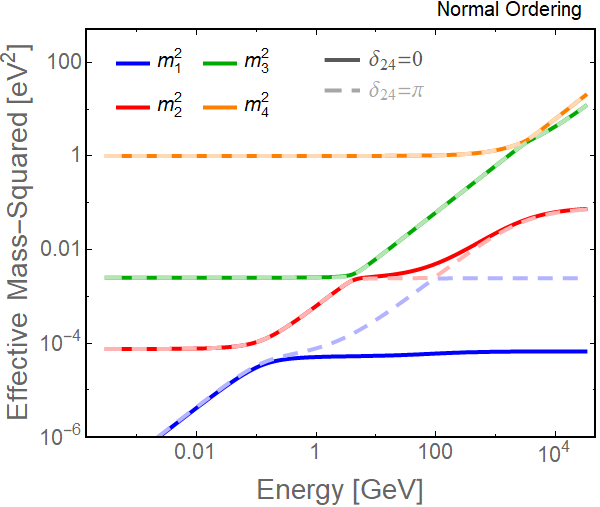}}
    \subfloat{\includegraphics[width=0.49\textwidth]{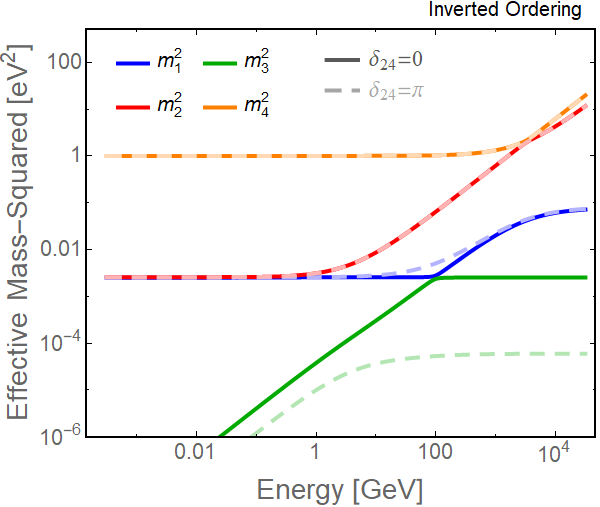}}
    \caption{Effective mass-squared values, representing the eigenvalues of Eq.~(\ref{eq:Ham}) for neutrinos in both normal (left) and inverted (right) orderings, as a function of neutrino energy. Three models are shown: In the upper panels, the standard picture with three active neutrinos (3$\nu$), is compared to a model with one light sterile neutrino where the CP violating phase $\delta_{24}$ is set to $0$. In the lower panels, two sterile neutrino models are compared with $\delta_{24}$ set to either $0$ or $\pi$. The absolute mass scale has been chosen so that the lightest neutrino is massless in vacuum. The oscillation parameters were set to $\Delta{m^2_{21}}=7.5\times10^{-5}~\mathrm{eV}^2$, $|\Delta{m^2_{31}}|=2.5\times10^{-3}~\mathrm{eV}^2$, $|\Delta{m^2_{41}}|=1~\mathrm{eV}^2$, $s_{12}^2=0.3$, $s_{13}^2=0.02$, $s_{23}^2=0.57$, $s_{14}^2=0.01$, $s_{24}^2=s_{34}^2=0.04$, and $\delta_{13}=\delta_{14}=0$. The matter density is set to 8.5~g/cm$^3$ with a ratio $N_n/N_e=1.08$.}
    \label{fig:massres}
\end{figure}

\begin{figure}
    \centering
    \subfloat{\includegraphics[width=0.49\textwidth]{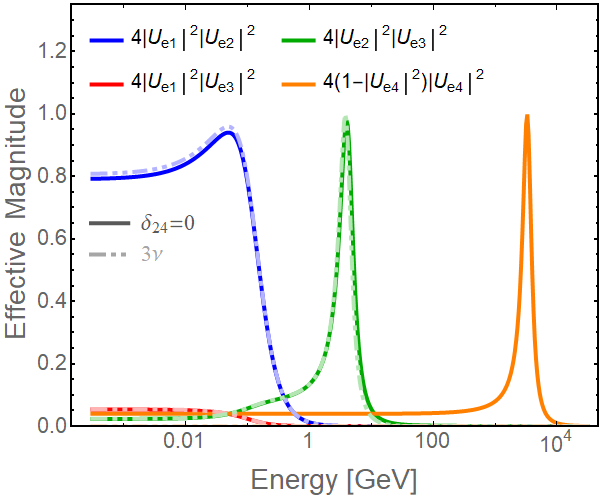}}
    \subfloat{\includegraphics[width=0.49\textwidth]{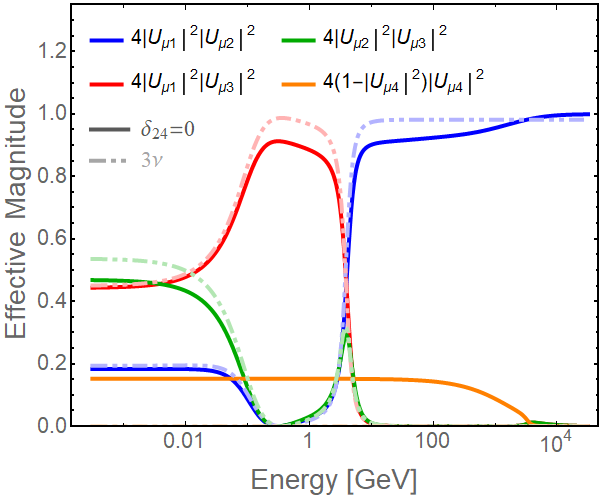}}\\
    \subfloat{\includegraphics[width=0.49\textwidth]{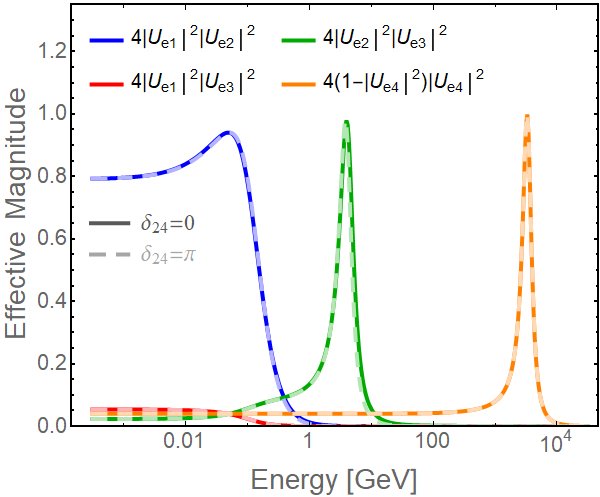}}
    \subfloat{\includegraphics[width=0.49\textwidth]{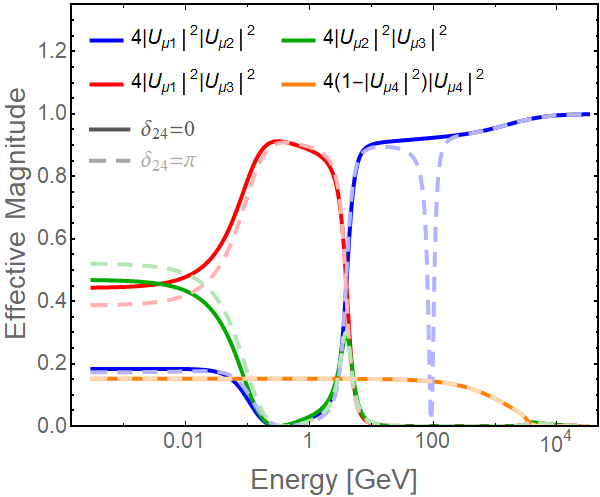}}
    \caption{Effective oscillation magnitudes associated with $\Delta_{21}$, $\Delta_{31}$, $\Delta_{32}$, and $\Delta_{4k}$ (see Equation~\ref{eq:probVac}) as a function of neutrino energy. The latter is taken as a combination of all three mass-squared difference terms involving the fourth mass state, which are approximately of equal frequency at this scale. Left: Magnitudes associated with $\nu_e$ disappearance probabilities. Right: Magnitudes associated with $\nu_\mu$ disappearance probabilities. Top: Comparison between 3$\nu$ and a sterile neutrino scenario with $\delta_{24}=0$. Bottom: Comparison between sterile neutrino scenarios with $\delta_{24}$ set to either $0$ or $\pi$. All plots apply to neutrinos in normal ordering. The same parameters as in Fig.~\ref{fig:massres} were used.}
    \label{fig:mixres}
\end{figure}

The general solutions to the Hamiltonian in Eq. \ref{eq:Ham} have a rich phenomenology that is difficult to express in analytical form. For the purposes of this analysis, Eq.~(\ref{eq:Ham}) is solved numerically in its full form using the software package OscProb~\cite{OscProb}. Fig.~\ref{fig:massres} shows an example of the impact of the matter potential on the effective values of the squared masses (eigenvalues of Eq.~(\ref{eq:Ham})) as a function of energy, assuming a medium of constant density for illustration purposes. Four resonances can be identified in the sterile neutrino models as regions of minimal distance between consecutive masses: one related to each pair ($s_{1k}$ \footnote{$s_{jk}$, $c_{jk}$ represent $\sin\theta_{jk}$ and $\cos\theta_{jk}$ respectively.}, $\Delta{m^2_{k1}}$) (at $\sim$0.05~GeV, 4~GeV and 3~TeV), and a second-order resonance connecting $s_{23}$, $s_{24}$, $s_{34}$, and $\Delta{m^2_{31}}$ (at $\sim$100~GeV), and with a strong dependence on $\delta_{24}$ as explained in section \ref{sec:high-E}.
The eigenvectors of Eq.~(\ref{eq:Ham}) define an effective mixing matrix which, in a medium of constant density, can be used to compute oscillation probabilities by direct replacement in the vacuum oscillation formula:
\begin{equation}
    \label{eq:probVac}
    \begin{array}{rl}
    P_{\alpha\beta}=\delta_{\alpha\beta}&-\ \sum_{j>k}4\,\mathrm{Re}[U_{\alpha{j}}U^*_{\beta{j}}U^*_{\alpha{k}}U_{\beta{k}}]\,\sin^2\frac{\Delta_{jk}L}{2}\\
    &-\ \sum_{j>k}2\,\mathrm{Im}[U_{\alpha{j}}U^*_{\beta{j}}U^*_{\alpha{k}}U_{\beta{k}}]\,\sin\Delta_{jk}L,
    \end{array}
\end{equation}
\noindent where $\Delta_{jk}=\Delta{m^2_{jk}}/2E$. Fig.~\ref{fig:mixres} shows examples of effective values of the magnitude of some terms from Eq.~(\ref{eq:probVac}) as a function of neutrino energy. The impact of the aforementioned resonances can be readily identified.
While the full numerical solutions exemplified above can already provide some insight, some exploration of common analytical approximations can be enlightening even if not used in the analysis. They are described in the following subsections.

\subsection{Large $|\Delta{m^2_{41}}|$ limit}

Anomalous oscillation results, such as LSND and MiniBooNE, are commonly interpreted as oscillations in a higher frequency than the solar and atmospheric scales. Under this scenario, the limit $\Delta{m^2_{41}}\rightarrow\infty$ can be considered in which all oscillations driven by $\Delta{m^2_{41}}$ are averaged out and observable only through scaling factors. Hereafter, this will be referred to as the high frequency (HF) region.
\\
Following Ref.~\cite{Maltoni:2007zf}, the mixing matrix $U$ can be split such that $U = \UNP \USM$, with $\USM = R_{23} \tilde{R}_{13} R_{12}$ containing only the active-active mixing elements, and $\UNP = R_{34} \tilde{R}_{24} \tilde{R}_{14}$ representing the active-sterile mixing. If the Hamiltonian is rotated with $\UNP$, it becomes approximately block-diagonal in the limit where $\Delta{m^2_{41}}\rightarrow\infty$:
\begin{equation}
\begin{array}{rl}
    \tilde{H} = \USM H_0 (\USM)^\dagger + (\UNP)^\dagger V \UNP 
    \approx \left(\begin{array}{cc}
        \tilde{H}^{(3)}  & 0 \\
        0  & \Delta_{41}
    \end{array}\right).
\end{array}
\end{equation}
The evolution matrix can then be expressed as:
\begin{equation}
    \label{eq:evol}
    S \approx \UNP \left(\begin{array}{cc}
        e^{-i\tilde{H}^{(3)}L} & 0 \\
        0 & e^{-i\Delta_{41}L}
    \end{array}\right) (\UNP)^\dagger .
\end{equation}
The remaining problem lies in the diagonalisation of $\tilde{H}^{(3)}$. For that, further approximations, which are valid in specific energy regimes, are employed. In general, a scale $\epsilon$ will be used to represent small quantities. The mixing parameters $s_{34}$, $s_{24}$, $s_{14}$, and $s_{13}$ will all be considered of $\mathcal{O}(\epsilon)$. Additionally, $\Delta{m^2_{21}}/\Delta{m^2_{31}}\sim s_{13}^2$ will be treated as $\mathcal{O}(\epsilon^2)$. In this approximation, probabilities can be written to $\mathcal{O}(\epsilon^2)$ as:
\begin{equation}
    P_{ee} \approx P_{ee}^{(3)} \cos2\theta_{14},
\end{equation}
\begin{equation}
    P_{e\mu} \approx c_{14}^2c_{24}^2P^{(3)}_{e\mu}+2c_{14}^2\mathrm{Re}[\UNP_{\mu2}{\UNP_{\mu1}}^*S^{(3)}_{e\mu}{S^{(3)}_{ee}}^*],
\end{equation}
\begin{equation}
    P_{\mu e} \approx c_{14}^2c_{24}^2P^{(3)}_{\mu e}+2c_{14}^2\mathrm{Re}[\UNP_{\mu2}{\UNP_{\mu1}}^*S^{(3)}_{\mu e}{S^{(3)}_{ee}}^*],
\end{equation}
\begin{equation}
    P_{\mu\mu} \approx P_{\mu\mu}^{(3)} \cos2\theta_{24} + 2c_{24}^2\mathrm{Re}[\UNP_{\mu2}\UNP_{\mu1}{}^*S^{(3)}_{\mu\mu}({S^{(3)}_{e\mu}}^*+{S^{(3)}_{\mu e}}^*)],
\end{equation}
\noindent where $\UNP_{\mu1}=-s_{14}s_{24}e^{i\delta_{14}-i\delta_{24}}$, $\UNP_{\mu2}=c_{24}$, and $S_{\alpha\beta}$ correspond to elements of the evolution matrix in Eq.~\ref{eq:evol}. The effect of mixing with sterile neutrinos is given by a scaling of the 3-neutrino submatrix probabilities. Additionally, some interference terms appear if both $s_{14}$ and $s_{24}$ are non-zero.

\subsubsection{The ORCA low energy regime}

The ORCA detector is most sensitive to neutrinos in the energy range of $3-100$ GeV considered in this analysis, crossing the Earth with paths of mean density varying between 3 and 9 g/cm$^3$. In the lower part of this energy range ($E < 10$ GeV), when $V_e=\sqrt{2}G_FN_e\sim\Delta_{31}=\Delta{m^2_{31}}/2E$, to leading order in small quantities, $\tilde{H}^{(3)}$ simplifies to:
\begin{equation}
    \tilde{H}^{(3)} \approx R_{23} \left(\begin{array}{ccc}
        V_e & 0 & \Delta_{31}c_{13}s_{13}e^{-i\delta_{13}} \\
        0 & 0 & 0 \\
        \Delta_{31}c_{13}s_{13}e^{i\delta_{13}} & 0 & \Delta_{31}c_{13}^2
    \end{array}\right) R_{23}^\dagger .
\end{equation}
This approximately 2-flavour form can be readily solved leading to the well-known MSW resonance of $\theta_{13}$:
\begin{equation}
    \tilde{H}^{(3)} \approx R_{23} \tilde{R}_{13}^m.
    \left(\begin{array}{ccc}
        -\Delta_{31}^m/2 & 0 & 0 \\
        0 & 0 & 0 \\
        0 & 0 & \Delta_{31}^m/2
    \end{array}\right) (\tilde{R}_{13}^m)^\dagger R_{23}^\dagger+\mathrm{const.},
\end{equation}
\begin{equation}
    \Delta_{31}^m = \sqrt{(\Delta_{31}\cos2\theta_{13}-V_e)^2+\Delta_{31}^2\sin^22\theta_{13}},
\end{equation}
\begin{equation}
    \sin2\theta_{13}^m = \frac{|\Delta_{31}|}{\Delta_{13}^m}\sin2\theta_{13},
\end{equation}
\noindent where $\tilde{R}_{13}^m$ represents the effective generalised unitary rotation matrix in the $1\mbox{-}3$ plane, parametrised by the effective mixing angle $\theta_{13}^m$ and the unchanged phase $\delta_{13}$.

All effects arising from the presence of sterile neutrino are constrained to vacuum-like mixing through $\UNP$ as in Eq.~(\ref{eq:evol}).

\subsubsection{The ORCA high energy regime}
\label{sec:high-E}
\begin{figure}
    \centering
    \subfloat{\includegraphics[width=0.49\textwidth]{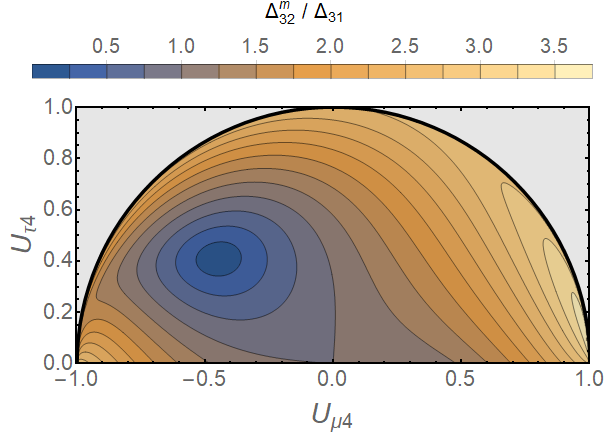}}
    \subfloat{\includegraphics[width=0.49\textwidth]{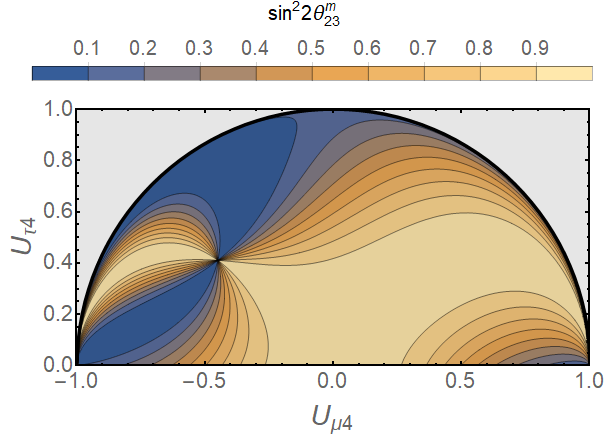}}
    \caption{Effective parameters $\Delta_{32}^m/\Delta_{31}$ (left) and $\sin^22\theta_{23}^m$ (right) as a function of real values of $U_{\mu4}$ and $U_{\tau4}$, for a neutrino energy of 20 GeV. Here, $\Delta{m^2_{31}}=2.5\times10^{-3}~\mathrm{eV}^2$, $s_{23}^2=0.57$, and a matter density of 8.5~g/cm$^3$ with a ratio $N_n/N_e=1.08$ were assumed. The resonance and antiresonance described in Equations (\ref{eq:vnres}) and (\ref{eq:antires}) are visible on the right as regions of maximum and minimum $\sin^22\theta_{23}^m$. At the point where they seem to meet, a pole exists where $\Delta_{32}^m\rightarrow0$ and $\sin^22\theta_{23}^m$ becomes undefined.}
    \label{fig:23res}
\end{figure}

At higher energies ($E\gtrsim10$~GeV), the matter potential starts to dominate. However, a new resonance can still be found when $\Delta_{31}/V_n$ is of $\mathcal{O}(\epsilon^2)$. In this regime, $\tilde{H}^{(3)}$ is expressed in leading order as:
\begin{equation}
    \tilde{H}^{(3)} \approx 
    \left(\begin{array}{ccc}
        V_e & 0 & 0 \\
        0 & \Delta_{31}s_{23}^2+V_n |\UNP_{s2}|^2 & \Delta_{31}s_{23}c_{23}+V_n {\UNP_{s2}}^*\UNP_{s3} \\
        0 & \Delta_{31}s_{23}c_{23}+V_n \UNP_{s2}\UNP_{s3} & \Delta_{31}c_{23}^2+V_n|\UNP_{s3}|^2
    \end{array}\right) ,
\end{equation}
where $\UNP_{s2}=-c_{34}s_{24}e^{i\delta_{24}}$ and $\UNP_{s3}=-s_{34}$. Once again, the Hamiltonian is approximately block diagonal and can be easily solved to give:
\begin{equation}
    \tilde{H}^{(3)} \approx R_{23}^m
    \left(\begin{array}{ccc}
        V_e & 0 & 0 \\
        0 & -\Delta_{32}^m/2 & 0 \\
        0 & 0 & \Delta_{32}^m/2
    \end{array}\right) (R_{23}^m)^\dagger+\mathrm{const.},
\end{equation}
\begin{equation}
    \begin{array}{rl}
    {\Delta_{32}^m}^2 = &[\Delta_{31}\cos2\theta_{23}+(|\UNP_{s3}|^2-|\UNP_{s2}|^2)V_n]^2\ +\\
    &|\Delta_{31}\sin2\theta_{23}+2V_n {\UNP_{s2}}^*\UNP_{s3}|^2 ,
    \end{array}
\end{equation}
\begin{equation}
    \sin2\theta_{23}^m = \frac{1}{\Delta_{32}^m}|\Delta_{31}\sin2\theta_{23}+2V_n{\UNP_{s2}}^*\UNP_{s3}| .
\end{equation}
This new resonance corresponds to a second order effect that couples the $2\mathrm{-}3$ sector indirectly via $s_{24}$ and $s_{34}$. It provides a very rich structure having two main features: a resonance when $\sin2\theta_{23}^m\rightarrow1$ and an antiresonance when $\sin2\theta_{23}^m\rightarrow0$ at finite $V_n$. The resonance conditions are:
\begin{equation}
    \label{eq:vnres}
    V_n = \frac{\Delta_{31}\cos2\theta_{23}}{(|\UNP_{s2}|^2-|\UNP_{s3}|^2)} \Rightarrow  \sin2\theta_{23}^m=1 ,
\end{equation}
\begin{equation}
    \label{eq:antires}
    V_n = -\frac{\Delta_{31}\sin2\theta_{23}}{2{\UNP_{s2}}^*\UNP_{s3}} \Rightarrow  \sin2\theta_{23}^m=0 .
\end{equation}
A pole exists when both conditions are satisfied, as $\Delta_{32}^m\rightarrow0$ and no mixing is possible. The structure of these resonances is shown in Fig.~\ref{fig:23res}.
Since $\theta_{23}$ is close to maximal, the antiresonance of Eq.~(\ref{eq:antires}) is the most noticeable effect in this regime. The antiresonance occurs for neutrinos when $\cos\delta_{24}\Delta{m^2_{31}}<0$ or for antineutrinos when $\cos\delta_{24}\Delta{m^2_{31}}>0$, and is only exact for $\delta_{24}=0\mbox{ or }\pi$. Hence, there is a degeneracy between mass ordering and $\mathrm{sign}(\cos\delta_{24})$, enhanced by the maximal value of $\sin2\theta_{23}$, which suppresses NMO contributions from the resonance term in Eq.~(\ref{eq:vnres}).

\subsection{Finite $|\Delta{m^2_{41}}|$ regime}

At values of $\Delta{m^2_{41}}$ for which the associated oscillations cannot be averaged out, no simplifying approximations are known to us at the time of writing. In ORCA, this corresponds to values of $\Delta{m^2_{41}} \lesssim 0.1$ eV$^2$, this regime will be referred to as the low frequency (LF) region. In this case, many interference terms are present and the probability formulas can become exceedingly complex. Nevertheless, a full numerical solution is possible on all regimes considered in the analysis, and it is used to extend the results through six orders of magnitude in $\Delta{m^2_{41}}$. For simplicity, $\Delta{m^2_{41}}$ will be restricted to positive values.

\section{Sterile Neutrino Analysis}\label{sec:analysismethod}
The analysis presented here is based on detailed Monte Carlo (MC) simulations as described in Ref.~\cite{ORCA_NMO_Paper}. Neutrino interactions are generated with gSeaGen~\cite{gSeaGen}, which is based on GENIE~\cite{GENIE}. Secondary particles and their emitted Cherenkov light are propagated with KM3Sim~\cite{KM3Sim}, a software package based on GEANT4~\cite{GEANT4}. The atmospheric neutrino flux is computed from the Honda model~\cite{Honda} for the Gran Sasso site without mountain over the detector, assuming minimum solar activity. Atmospheric muons are generated with MUPAGE~\cite{MUPAGE, MUPAGE2}, and propagated with KM3~\cite{KM3}.
\\
Event reconstruction is performed via a maximum likelihood fit to shower and track hypotheses. Background events arising from noise and atmospheric muons are rejected with two independent Random Decision Forests (RDF) trained on MC simulations. A third RDF was used to separate neutrino candidates into three topology classes defined by the output score of the RDF, trained to identify track-like events. Events with a track score larger than 0.7 are labelled as track-like, track scores less than 0.3 are labelled as shower-like, and other values are labelled as an intermediate topology. Moreover, as in Ref.~\cite{ORCA_NMO_Paper}, only upgoing events are considered in order to get rid of the atmospheric muon contamination.
\\
Instead of using parametrised response functions as in Ref.~\cite{ORCA_NMO_Paper}, the analysis reported here is based on the aforementioned MC simulations to directly model the detector response. The two approaches have been compared and found consistent.
\\
The MC-based modelling of the detector response is implemented in the KM3NeT framework Swim \cite{swim}. The detector response is represented by a 4-dimensional matrix, as a function of  true and reconstructed neutrino energy $E, \, E'$, and zenith angle $\theta, \, \theta'$, for each interaction channel $\nu_x$, $R^{[\nu_x \rightarrow i]}(E, \theta, E', \theta')$. Each entry of this matrix summarises in a single dimensionless coefficient the efficiency of detection, classification and probability of reconstruction for a given true bin $(E,\theta)$. Therefore, $R$ incorporates all the effects related both to the detector and to the event selection. More details on this approach, can be found in Ref.~\cite{swim}. The binning scheme, for the detector response matrix, used in this analysis is shown in Tab.\ \ref{tab:binsSwim}.\ Since the atmospheric neutrino flux follows a power law in energy, equal-width bins in $\log_{10}E$ are chosen. The same choice is adopted for reconstructed events histograms, as the relative energy resolution in ORCA is, to first order, constant above $\sim 10$ GeV, $\delta E / E \simeq \delta(\log_{10} E) \simeq 15\%$ \cite{intrinsic_limits}. 
\\
A binning of constant width in $\cos\theta_Z$ is used. This is motivated by the fact that the solid angle covered by an interval of zenith angle $\theta_1 \leq \theta \leq \theta_2$ is proportional to $|\cos\theta_1 - \cos\theta_2|$ and, considering to first order the atmospheric neutrino flux as isotropic, this choice yields equally populated bins along the zenith angle axis. 
\begin{table}[h]
	\centering
	\begin{tabular}{|c|c|c|c|c|}
		\hline
		& $E$ [GeV]& $\cos\theta_Z$ & $E'$ [GeV]& $\cos\theta_Z'$ \\
		\hline
		Bins & 40 & 40 &  20 & 20 \\
		\hline
		Range & $[1, 100]$ & $[-1, 0]$ & $[2, 100]$ & $[-1, 0]$ \\
		\hline
	\end{tabular}
	\caption{Bin choice for the MC-based response matrix, $R$, used in this analysis. Energy bins are in $\log_{10}$ space.}
	\label{tab:binsSwim}
\end{table}
\\
For reconstructed event histograms, the choice of binning granularity is dominated by the detector resolutions. The bin width should be comparable with the typical error on the reconstructed variable. Moreover, it should account for a sufficiently smooth sampling of the detector response, to minimise the finite MC statistics issues, which can result in overestimations of sensitivity \cite{swim}. Statistical fluctuations due to the sparse MC effect are taken into account by following the “Beeston and Barlow method" \cite{beeston}. 
\\
The values of the standard neutrino parameters used in this analysis is taken from the NuFit v4.1 global fit result with Super-Kamiokande (SK) data~\cite{NuFit41} and summarised in Tab.~\ref{tab:benchmarkoscparam}, for both normal (NO) and inverted ordering (IO). Current fits have large errors on $\delta_{CP}$. The impact of such variable in the analysis has been tested and found to be negligible. For this reason, its value is fixed to the ones reported in Tab.~\ref{tab:benchmarkoscparam}. Moreover, $\dmf > 0$ is always assumed. Oscillation probabilities are evaluated with the software package OscProb \cite{OscProb}, and to account for Earth's matter effects the PREM model \cite{PREM} with 44 layers is used.
\begin{table}[h]
	\centering
	\begin{tabular}{|c|c|c|c|c|c|c|}
		\hline
		& $\sin^2 \theta_{12}$ & $\sin^2\theta_{23}$ & $\sin^2\theta_{13}$ & $\dcp$ & $\Delta m_{21}^2 (\rm eV^2)$ & $\Delta m_{31}^2 (\rm eV^2)$ \\
		\hline
		NO & 0.310 & 0.563 & 0.02237 & $221^\circ$ & 7.39 $\times$ $10^{-5}$ & 2.528 $\times$ $10^{-3}$ \\
		\hline
		IO & 0.310 & 0.565 & 0.02259 & $282^\circ$ & 7.39 $\times$ $10^{-5}$ & $-2.510 \times$ $10^{-3}$ \\
		\hline
	\end{tabular}
	\caption{Benchmark oscillation parameters for NO and IO, taken from the NuFit v4.1 result \cite{NuFit41}.}
	\label{tab:benchmarkoscparam}
\end{table}
\\
The above information can be used to define the distinguishability $S_\sigma$, as a quick estimator of sensitivity of measurements, with the goal of illustrating the impact of a sterile neutrino in the event distributions, as
\begin{equation}
S_{\sigma} = \frac{(N_{\rm Sterile} - N_{\rm Standard}) |N_{\rm Sterile} - N_{\rm Standard}|}{N_{\rm Sterile} },
\end{equation}
where $N_{\rm Sterile}$ and $N_{\rm Standard}$ are the number of events, as a function of reconstructed energy and zenith angle, in the sterile and standard hypothesis respectively.
\begin{figure}
	\centering
	\includegraphics[width=1.\linewidth]{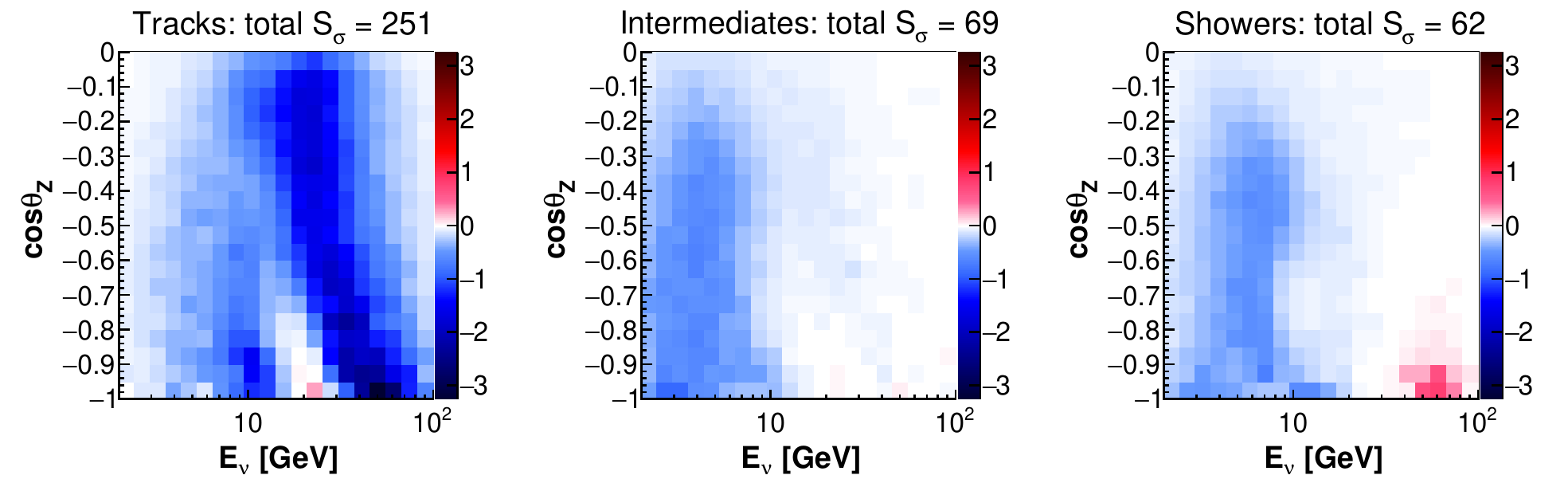}
	\caption{$S_\sigma$ distribution of the three topologies considered in the analysis (tracks, intermediates and showers) assuming three years of data taking. The colour scale denotes the $S_\sigma$ value for each bin, whereas the total $S_\sigma$ is reported on top of the plots: the high value obtained is due to the normalisation. The sterile neutrino parameters are $\sin^2\theta_{14} = 0$, $\sin^2\theta_{24} = 0.03$,  $\sin^2\theta_{34} = 0.05$, $\Delta m_{41}^2 = 1 \, \eV$.}
	\label{fig:chi2paperstd3years}
\end{figure}
\begin{figure}
	\centering
	\includegraphics[width=1.\linewidth]{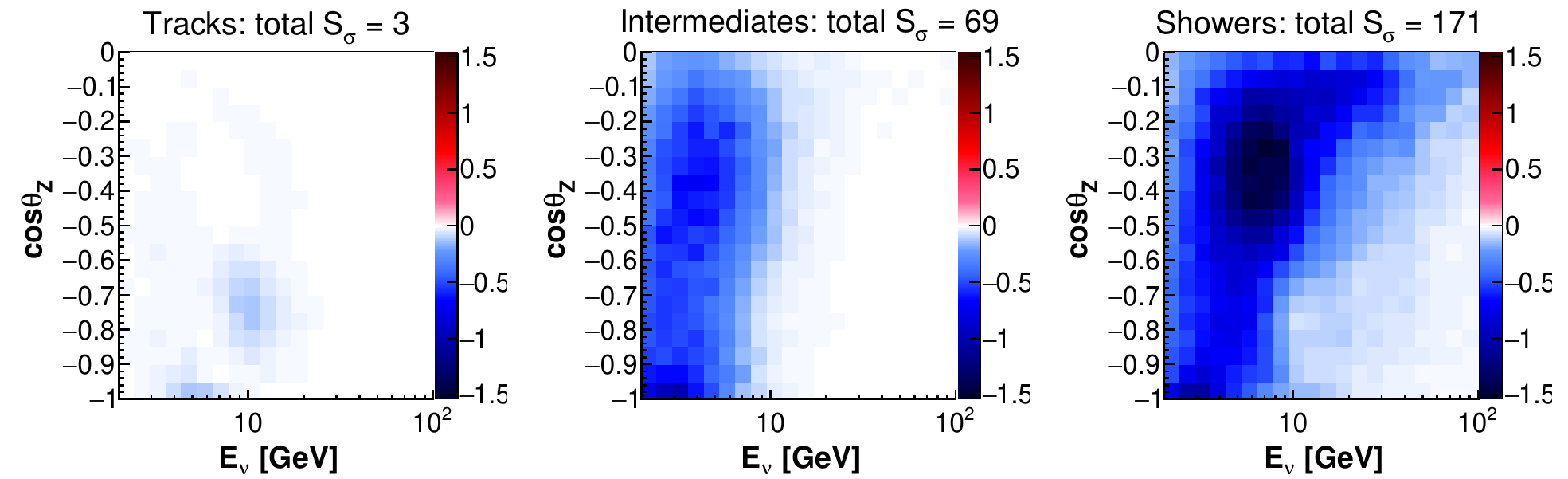}
	\caption{$S_\sigma$ distribution of the three topologies considered in the analysis (tracks, intermediates and showers) assuming three years of data taking. The colour scale denotes the $S_\sigma$ value for each bin, whereas the total $S_\sigma$ is reported on top of the plots: the high value obtained is due to the normalisation. The sterile neutrino parameters are $\sin^2\theta_{14} = 0.05$,  $\sin^2\theta_{24} = \sin^2\theta_{34} = 0$, $\Delta m_{41}^2 = 1 \, \eV$.}
	\label{fig:chi2paperue43years}
\end{figure}
\begin{figure}
	\centering
	\includegraphics[width=1.\linewidth]{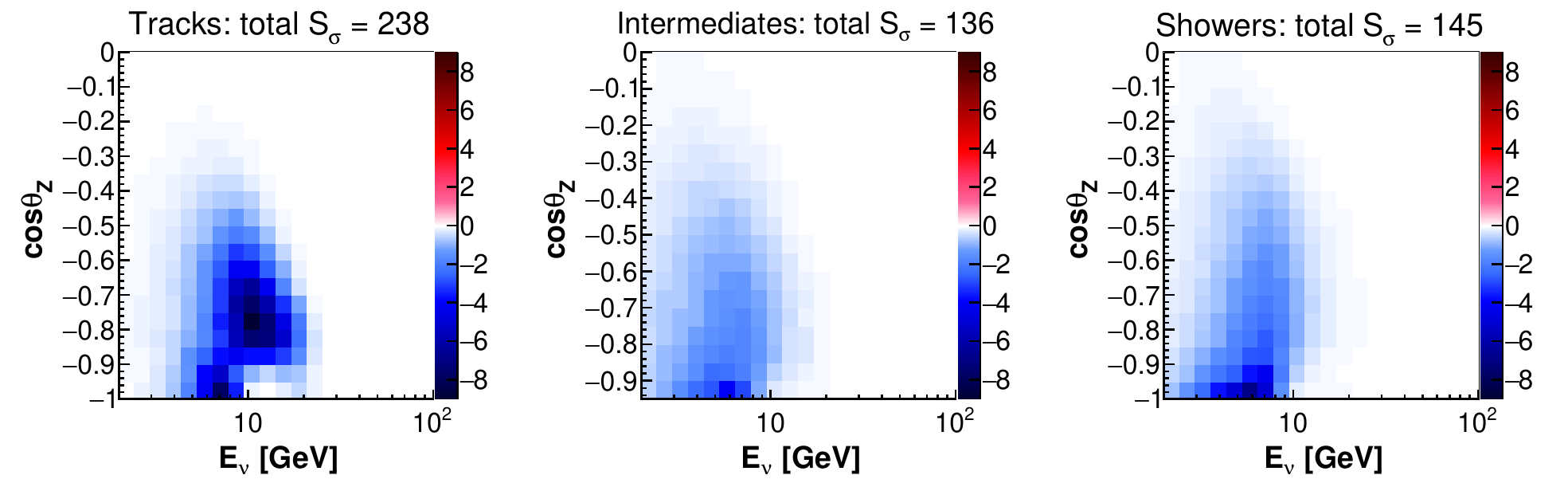}
	\caption{$S_\sigma$ distribution of the three topologies considered in the analysis (tracks, intermediates and showers) assuming three years of data taking. The colour scale denotes the $S_\sigma$ value for each bin, whereas the total $S_\sigma$ is reported on top of the plots: the high value obtained is due to the normalisation. The sterile neutrino parameters are $\sin^2\theta_{14} = \sin^2\theta_{24} = \sin^2\theta_{34} = 0.01$,  $\Delta m_{41}^2 = 10^{-4} \, \eV$.}
	\label{fig:chi2paperlowmass3years}
\end{figure}
\\
Fig. \ref{fig:chi2paperstd3years} shows the distinguishability distribution for a sterile neutrino in the HF region, for non-zero $\theta_{24}$ and $\theta_{34}$, assuming three years of ORCA data taking. The presence of the sterile neutrino mainly impacts the track-like events in the form of a deficit of upgoing events at higher energies ($E' \geq 40$ GeV). Therefore, this region of the sterile parameter space can be well constrained also by neutrino telescopes whose energy threshold is higher than that of ORCA, such as ANTARES \cite{ANTARES_Sterile} and IceCube/DeepCore \cite{DeepCore_Steriles}.
\\
To understand the sensitivity to $\theta_{14}$ in particular, the distinguishability for a sterile neutrino in the HF region and $\sin^2\theta_{14}=0.05$ is shown in Fig. \ref{fig:chi2paperue43years}. In this case, the shower-like events are the most affected and mainly for energies $< 20$ GeV. It follows that ORCA is well suited to test $\theta_{14}$.
\\
Finally, Fig. \ref{fig:chi2paperlowmass3years} shows the impact of a sterile neutrino with $\Delta m_{41}^2 = 10^{-4}$ eV$^2$. In this case, the energy region $E' < 10$ GeV is the most significant, and all the three event topologies are highly impacted. This applies also for $\Delta m_{41}^2 = 10^{-2}, 10^{-3}, 10^{-5} \, \eV$.
\\
The sensitivity evaluation is based on the minimisation of a negative log-likelihood function describing the agreement between a model prediction and observed data. This is done with the Asimov approach \cite{Asimov} assuming the negative log-likelihood follows a chi-squared distribution. Specifically, the negative log-likelihood function is defined as: 
\begin{equation}
\begin{split}
\chi^2 = & -2 \log L = \chi^2_{\rm stat} + \chi^2_{\rm syst} = \\
& 2 \sum_{i=1}^{N_{E'}} \sum_{j=1}^{N_{\cos \theta'}} \sum_{t=1}^{3} \left[ N_{ijt}^{\rm model}(\eta) - N_{ijt}^{\rm data}
+ N_{ijt}^{\rm data}\rm{log} \left( \frac {\it N_{\rm ijt}^{\rm data}}{\it N_{ijt}^{\rm model}(\eta)} \right) \right] \\
& + \sum_{k=1}^{N_{\rm Syst}} \left( \frac{\eta'_{k} - \braket{\eta'_{k}}}{\sigma_{\eta'_k}} \right)^2 ,
\end{split}
\end{equation}
where $N_{ijt}^{\rm model}$ and $N_{ijt}^{\rm data}$ represent the number of expected and measured events in bin ($i,j$) respectively and the sum over $t$ runs over the three event topologies: tracks, intermediates and showers. $\eta$ represents the model parameters, which comprise both the oscillation parameters listed in Tab.\ \ref{tab:benchmarkoscparam}, and nuisance parameters $\eta'$, which are related to systematic uncertainties. The second sum runs over the nuisance parameters and $\braket{\eta'_{k}}$ is the assumed prior of the parameter $k$ and $\sigma_{\eta'_k}$ its uncertainty. The set of free parameters considered in this analysis, together with the assumed gaussian priors with mean $\mu$ and standard deviation $\sigma$, is summarised in Tab.\ \ref{tab:systlist}. 
\begin{table} 
	\centering
	\begin{tabular}{|c |c|} 	
		\hline
		\bf{Parameter} & \bf{Gaussian Prior ($\mu \pm \sigma$)}  \\ [0.5ex] 
		\hline
		$\nu_e/\bar{\nu}_e$ & $0 \pm 0.07$ \\ 
		\hline
		$\nu_\mu/\bar{\nu}_\mu$ & $0 \pm 0.05$ \\ 
		\hline
		$\nu_e/\nu_\mu$ & $0 \pm 0.02$ \\ 
		\hline
		NC Scale & No prior \\
		\hline
		Energy Scale & $1 \pm 0.05$\\
		\hline
		Energy Slope & No prior \\
		\hline
		Zenith Angle Slope & $0 \pm 0.02$\\
		\hline
		Track Normalisation & No Prior \\
		\hline
		Intermediate Normalisation & No Prior \\
		\hline
		Shower Normalisation & No Prior \\
		\hline
		$\Delta m_{31}^2$ & No prior \\ 
		\hline
		$\theta_{13}$ & $\theta_{13} \pm 0.13^\circ$ \\
		\hline
		$\theta_{23}$ & No prior \\[1ex]
		\hline
	\end{tabular}
	\caption{List of fitted values and relative gaussian priors considered in this analysis. $\theta_{13}$ refers to the values listed in Tab.\ \ref{tab:benchmarkoscparam}}
	\label{tab:systlist}
\end{table} 
Where the uncertainties on the neutrino flux are taken from Ref. \cite{flux_syst} and the uncertainty on the detector energy scale follows the investigations reported in Ref. \cite{loi} (section 3.4.6). Specifically:
\begin{enumerate}
	\item \label{sysnew-2} the ratio between the total number of $\nu_e$ and $\bar{\nu}_e$ is allowed to vary with a standard deviation of 7$\%$ of the parameter's nominal value,
	\item the ratio between the total number of $\nu_\mu$ and $\bar{\nu}_\mu$ is allowed to vary with a standard deviation of 5$\%$ of the parameter's nominal value,
	\item the ratio between the total number of $\nu_e$ and $\nu_\mu$ is allowed to vary  with a standard deviation of 2$\%$ of the parameter's nominal value,
	\item the number of NC events is scaled by the \textit{NC scale} factor, to which no constraint is applied,
	\item the absolute \textit{energy scale} of the detector, which depends on the knowledge of the PMT efficiencies and the water optical properties, as discussed in Ref.~\cite{ORCA_NMO_Paper}, is allowed to vary with a standard deviation of 5\% around its nominal value,
	\item the \textit{energy slope} of the neutrino flux energy distribution is allowed to vary without constraint,
	\item \label{sysnew-1} the ratio of upgoing to horizontally-going neutrinos, the \textit{zenith angle slope}, is allowed to vary with a standard deviation of $2\%$ of the parameter's nominal value,
	\item  \label{sysnew-L} the number of events in the three classes is allowed to vary without constraints,
	\item $\Delta m_{31}^2$ and $\theta_{23}$ are allowed to vary without constraints,
	\item $\theta_{13}$ is allowed to vary within a $1\sigma$ window of the parameter's nominal value, which corresponds to $0.13^\circ$ for both NO and IO.
\end{enumerate}
In the following section, the ORCA sensitivity to the active-sterile parameters is presented.

\section{Sensitivity Results}
\label{sec:sensi}
The ORCA sensitivity to the active-sterile mixing angles is here presented. The Asimov dataset is obtained using the parameters in Tab. \ref{tab:benchmarkoscparam}, assuming no sterile neutrino in NO and IO. No assumption is made on NMO: the fit is marginalised over NMO. This allows to conservatively take into account degeneracies between NMO and the sterile parameters.
\\
At the SBL neutrino mass scale, $\dmf \sim 1$ $\eV$, correlated constraints in the $\theta_{24} -\theta_{34}$ parameter space are obtained. And, for a more general analysis, sensitivities to the mixing elements $\theta_{14}$, $\theta_{24}$, $\theta_{\mu e}$ and $\theta_{34}$ over the range $\dmf \in [10^{-5}, 10] \; \eV$ are presented.

\subsection{Sensitivity to $\theta_{24} -\theta_{34}$ in the large $\dmf$ limit}
As shown in Fig.\ \ref{fig:chi2paperstd3years}, in this sterile mass region, the track channel appears to be the most effective in constraining $\theta_{24}$ and $\theta_{34}$.
\\
As stated in Sec.\ \ref{sec:phenomeno}, %\ref{sec:LSMS}, 
$\delta_{24}$ highly impacts the analysis due to matter effects. Therefore, $\delta_{24}$ is kept free in the fit. Whereas, we  investigated the impact of $\theta_{14}$ and found it to be negligible, therefore $\theta_{14}$ and $\delta_{14}$ are fixed to zero in this part of the analysis.
%%%%%%%%%%%%%%%%%%%%%%%%%%%%%%%%%%%%%%%%%%%%%%%%%%%%%%%%%%%%%%%%%%%
\begin{figure}[h]
\centering
\includegraphics[width=1.01\textwidth]{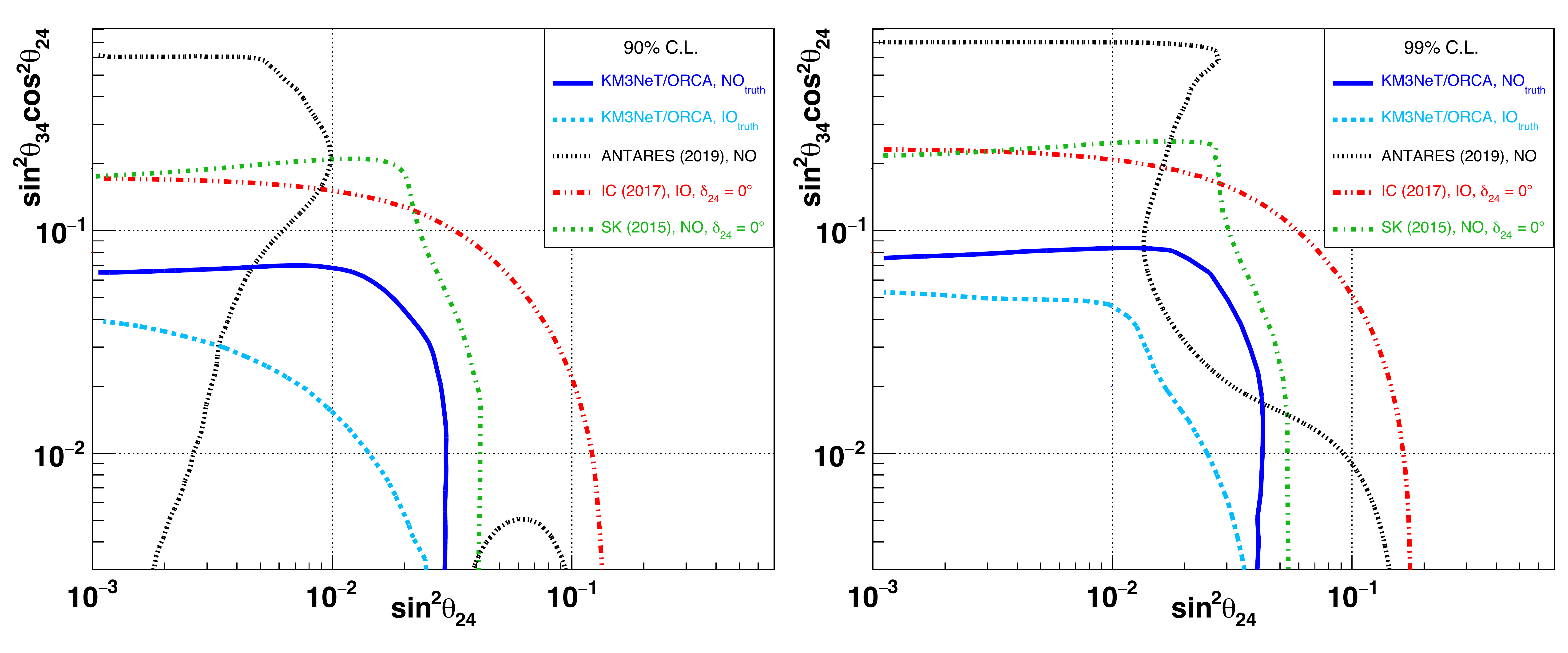}
\mycaption{The $90\%$ (left) and $99\%$ C.L. (right) KM3NeT/ORCA sensitivity to the mixing parameters $\theta_{24} - \theta_{34}$, with $\dmf = 1 \,\eV$, for three years of assumed data taking. The obtained sensitivity is compared with current upper limits from ANTARES \cite{ANTARES_Sterile}, IceCube/DeepCore (IC) \cite{DeepCore_Steriles} and SK \cite{SK}. If not explicitly stated, $\delta_{24}$ is free in the fit: this applies to the results from ORCA and ANTARES. The excluded region is the one on the top right of the lines.}
\label{fig:sen-Um-Ut}
\end{figure}
%%%%%%%%%%%%%%%%%%%%%%%%%%%%%%%%%%%%%%%%%%%%%%%%%%%%%%%%%%%%%%%%%%%%%
\\
Fig.\ \ref{fig:sen-Um-Ut} shows the 90$\%$ and 99$\%$ C.L. ORCA sensitivity on $\sin^2 \theta_{24}$ and $\sin^2 \theta_{34} \cos^2 \theta_{24}$ for three years of data taking. The ORCA sensitivity is compared to upper limits from other neutrino experiments, namely ANTARES \cite{ANTARES_Sterile}, IceCube/DeepCore \cite{DeepCore_Steriles} and SK \cite{SK}. In order to highlight the impact of $\delta_{24}$ in the final constraints, ANTARES has presented upper limits \cite{ANTARES_Sterile} with $\delta_{24}$ fixed to 0 and free. Allowing $\delta_{24}$ to be free worsens the constraints on $\theta_{24}$ and $\theta_{34}$ and it needs to be considered as a free parameter by all the analyses in which Earth matter effects are not negligible. Here, only the analysis with $\delta_{24}$ free is presented. The impact of this quantity in the ORCA sensitivity can be found in Ref. \cite{icrc_steriles}: it is maximal when $\sin^2 \theta_{24} = \sin^2 \theta_{34} \cos^2 \theta_{24}$, for which case it worsens the sensitivity by about a factor of two for $\sin^2 \theta_{24}$ and a factor three for $\sin^2 \theta_{34} \cos^2 \theta_{24}$. 
\\
Due to the degeneracy driven by NMO and $\delta_{24}$, discussed in Sec. \ref{sec:phenomeno}, the ORCA Asimov dataset in NO and $\delta_{24}$ free (blue line) can be directly compared with IceCube/DeepCore IO and $\delta_{24}=0$ (red line). For SK, upper limits with IO are not available, therefore the ones with NO and $\delta_{24}=0$ are here reported.
\\
From Fig. \ref{fig:sen-Um-Ut} it can be concluded that ORCA is competitive in constraining the mixing elements $\theta_{24}$ and $\theta_{34}$, and it is expected to improve the sensitivity to $\sin^2 \theta_{34} \cos^2 \theta_{24}$ by over a factor of two with respect to current limits.

\subsection{Sensitivity to $\theta_{24}$ for different $\dmf$ values}
\begin{figure}[h]
	\centering
	\includegraphics[width=1.01\textwidth]{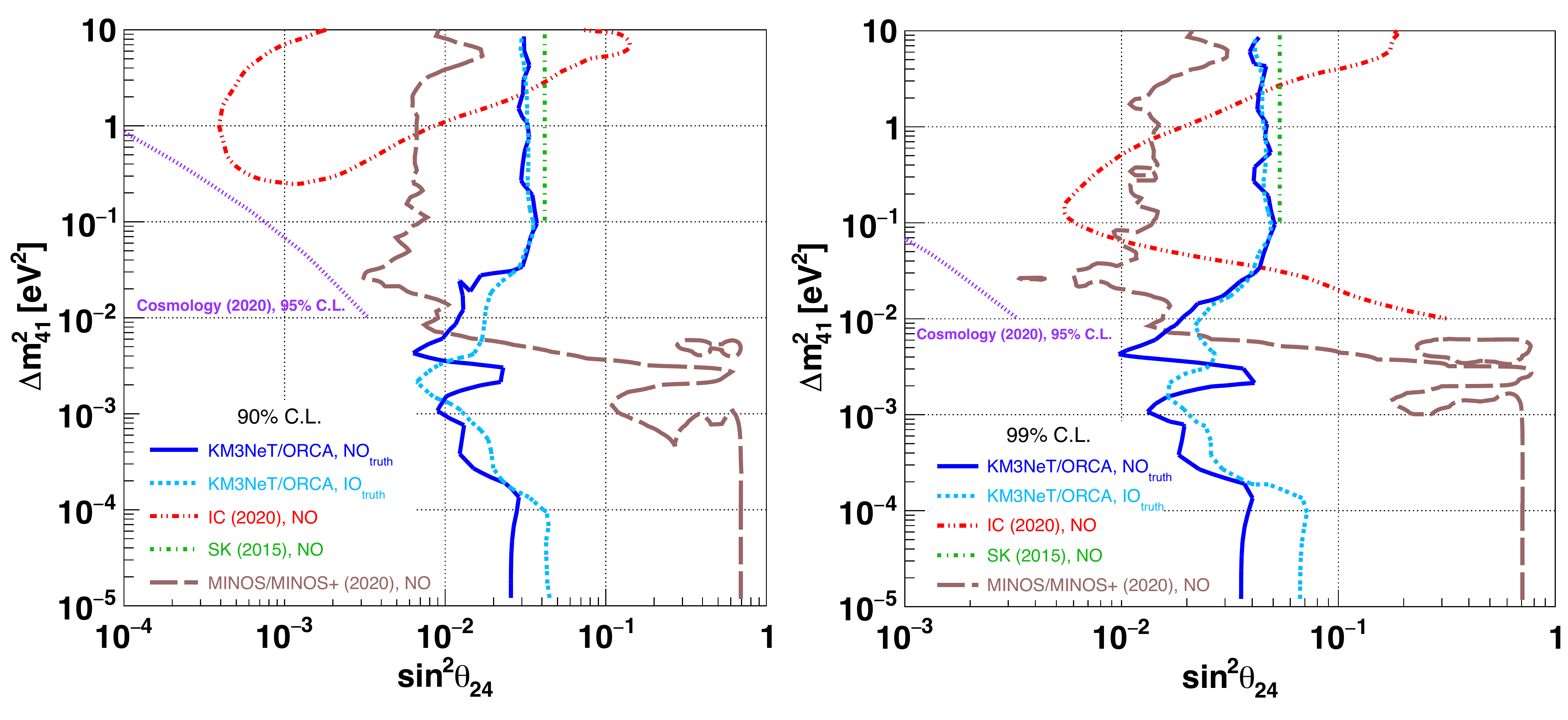}
	\mycaption{The $90\%$ (left) and $99\%$ C.L. (right) KM3NeT/ORCA sensitivity to the mixing parameter $\theta_{24}$, assuming three years of data taking. The obtained sensitivity is compared with current upper limits from cosmology \cite{strongboundCosmology}, MINOS/MINOS+ \cite{numu_disapp_minos}, IceCube (IC) \cite{IceCube_sterile} and SK \cite{SK}. The excluded region is the one on the right of the lines, for IceCube at $90\%$ C.L. it is the external region to the closed contour line.}
	\label{fig:sen-Um}
\end{figure}

Fig. \ref{fig:sen-Um} shows the $90\%$ and $99\%$ C.L. ORCA sensitivity to $\sin^2\theta_{24}$ assuming three years of data taking. For this analysis, $\theta_{14}$, $\theta_{34}$, $\delta_{14}$ and $\delta_{24}$ are set free in the fit, since their effects on the results of the analysis are expected to be not negligible. 
\\
The ORCA sensitivity is compared with upper limits from cosmology \cite{strongboundCosmology} for which only $95\%$ C.L. are available, and upper limits from MINOS/MINOS+ \cite{numu_disapp_minos}, IceCube \cite{IceCube_sterile} and SK \cite{SK}.
\\
Both plots show that ORCA is less competitive than MINOS/MINOS+ and IceCube for HF. KM3NeT/ARCA would be better suited to test $\sin^2\theta_{24}$ in this region. In the LF region, ORCA is able to improve current limits on $\sin^2\theta_{24}$ by more than one order of magnitude.
%%%%%%%%%%%%%%%%%%%%%%%%%%%%%%%%%%%%%%%%%%%%%%%%%%%%%%%%%%%%%%%%%%%%

\subsection{Sensitivity to $\theta_{14}$ for different $\dmf$ values}
\begin{figure}[H]
\centering
\includegraphics[width=0.55\textwidth]{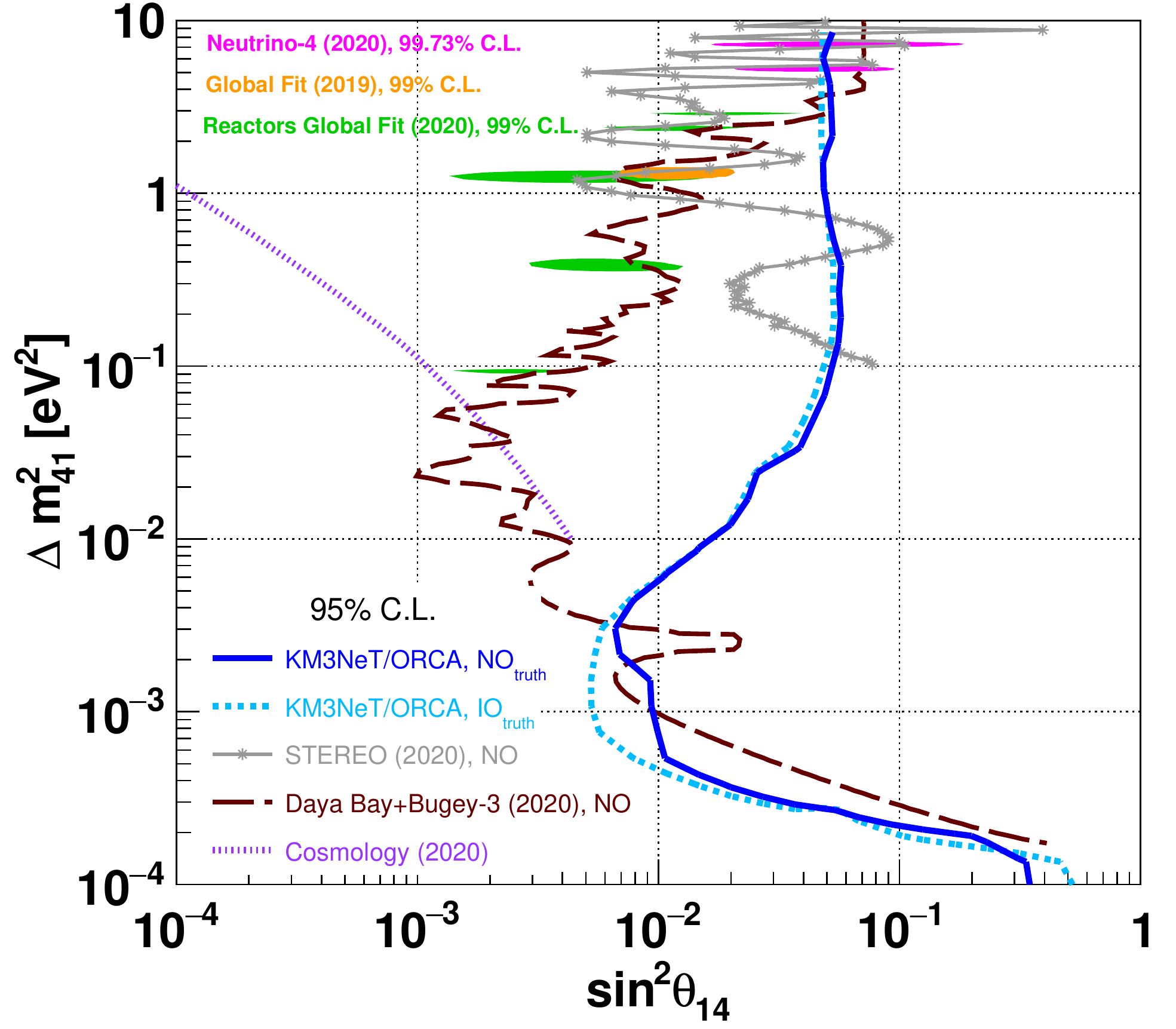}
\mycaption{The $95\%$ C.L. KM3NeT/ORCA sensitivity to the mixing parameter $\theta_{14}$, for different values of $\dmf$, for three years of data taking. Sensitivity results are compared with current upper limits from cosmology \cite{strongboundCosmology}, STEREO \cite{stereo_b}, and Daya Bay+Bugey-3 \cite{numu_disapp_minos}. Current anomaly regions are also reported, from Neutrino-4 \cite{neutrino-4}, global fits \cite{sterile_review} and reactors global fits \cite{globesfit}. The excluded region is the one on the right of the lines.}
\label{fig:sen-Ue}
\end{figure}

Fig. \ref{fig:sen-Ue} shows the $95\%$ C.L. ORCA sensitivity to $\sin^2\theta_{14}$ after three years of data taking. The choice to show the sensitivity at such a level of confidence is motivated by the goal to have a fair comparison with the other experiments, for which the majority of the available upper limits and sensitivity is reported at $95\%$ C.L. For this analysis, $\theta_{24}$, $\theta_{34}$, $\delta_{14}$ and $\delta_{24}$ are free in the fit, since their effects on the results of the analysis are expected to be not negligible.
Fig. \ref{fig:chi2paperue43years} shows that, in the HF region, shower-like events are the most affected by $\theta_{14}$ and in the optimal energy region for ORCA ($E' < 10$ GeV). However, they are concentrated in the nearly-horizontal region ($-0.1 <\cos\theta_{Z} < -0.6$). Nevertheless, ORCA has a competitive sensitivity to Daya Bay+Bugey-3 \cite{numu_disapp_minos} and STEREO \cite{stereo_b} in the HF region. Moreover, ORCA will also be able to test part of the Neutrino-4 allowed region \cite{neutrino-4}. On the contrary, the global fit regions can not be reached with three years of data taking.

%%%%%%%%%%%%%%%%%%%%%%%%%%%%%%%%%%%%%%%%%%%%%%%%%%%%%%%%%%%%%%%%%%%%%
\subsection{Sensitivity to $|U_{\mu e}|^2$ for different $\dmf$ values}
\begin{figure}[h]
	\centering
	\includegraphics[width=1\textwidth]{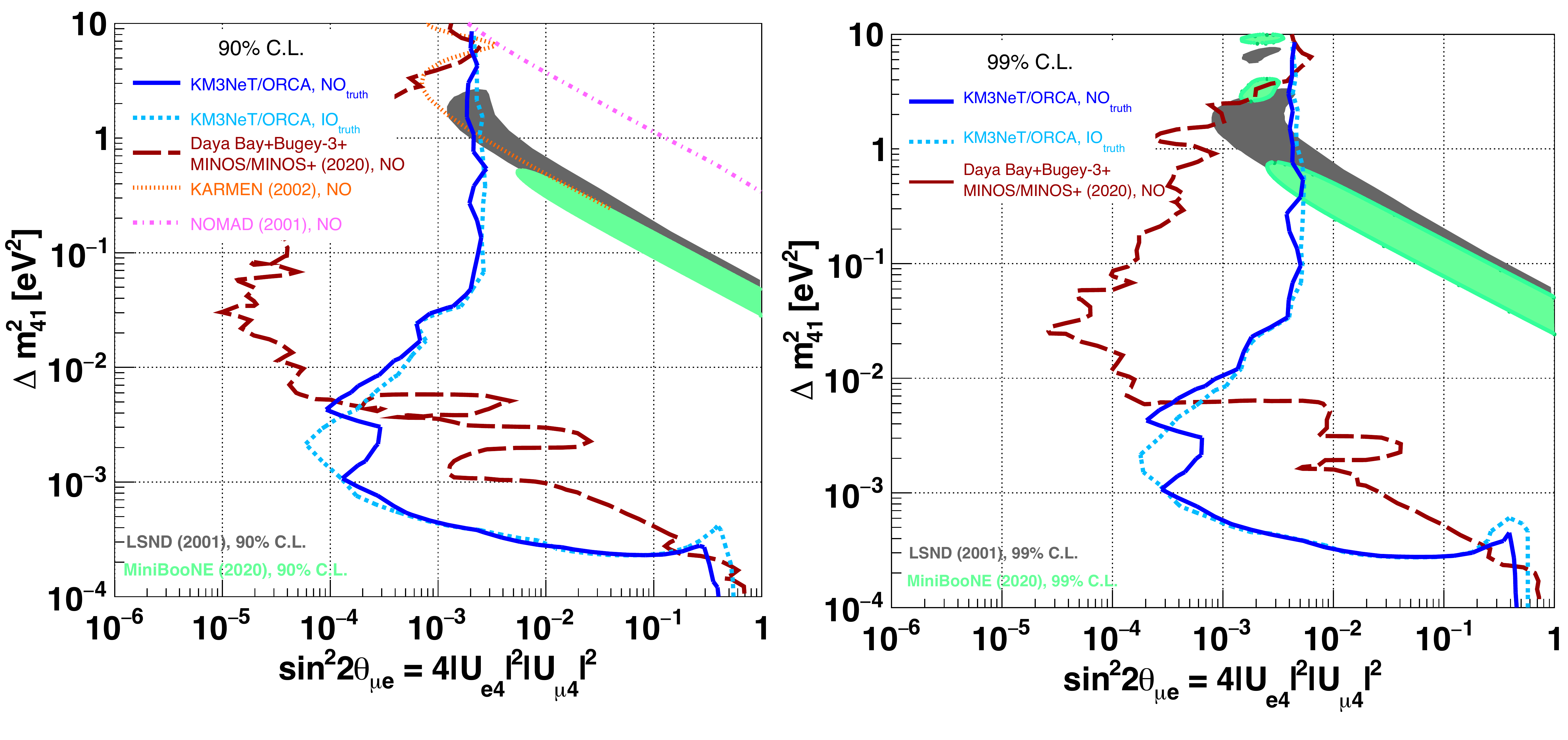}
	\mycaption{The $90\%$ (left) and $99\%$ C.L. (right) KM3NeT/ORCA sensitivity to the mixing parameter $|U_{\mu e}|^2$, assuming three years of data taking. Sensitivity results are compared with current upper limits from Daya Bay+Bugey-3+MINOS/MINOS+\cite{numu_disapp_minos}, KARMEN \cite{karmen} and NOMAD \cite{nomad}. Current anomaly regions from LSND \cite{LSND} and MiniBooNE \cite{Miniboone} are also reported. The excluded region is the one on the right of the lines.}
	\label{fig:sen-Umue}
\end{figure}

Since ORCA can observe both $\nu_e$ and $\nu_\mu$ disappearance, the effective mixing element $|U_{\mu e}|^2 = \sin^2 2\theta_{\mu e} = 4 |U_{e4}|^2|U_{\mu4}|^2$ can be constrained directly. In this case, $\theta_{14}$ and $\theta_{24}$ are left free in the fit, however, their combination is constrained to match the appropriate $\theta_{\mu e}$ value by introducing a penalty term in the likelihood with a very small prior uncertainty of $10^{-6}$. Fig. \ref{fig:sen-Umue} shows the $90\%$ and $99\%$ C.L. ORCA sensitivity to $|U_{\mu e}|^2$, compared with current upper limits from Daya Bay+Bugey-3+MINOS/MINOS+ \cite{numu_disapp_minos}, KARMEN \cite{karmen}, and NOMAD \cite{nomad}.
\\
Fig.\ \ref{fig:sen-Umue} shows that, after three years of data taking, ORCA will be able to test the majority of the LSND \cite{LSND} and MiniBoone \cite{Miniboone} anomaly region. Moreover, current limits on $\sin^2 2\theta_{\mu e}$ will be improved by 1-2 orders of magnitude in the LF region. 

%%%%%%%%%%%%%%%%%%%%%%%%%%%%%%%%%%%%%%%%%%%%%%%%%%%%%%%%%%%%%%%%%%%

\subsection{Sensitivity to $\theta_{34}$ for different $\dmf$ values}
\begin{figure}[h!]
\centering
\includegraphics[width=0.55\textwidth]{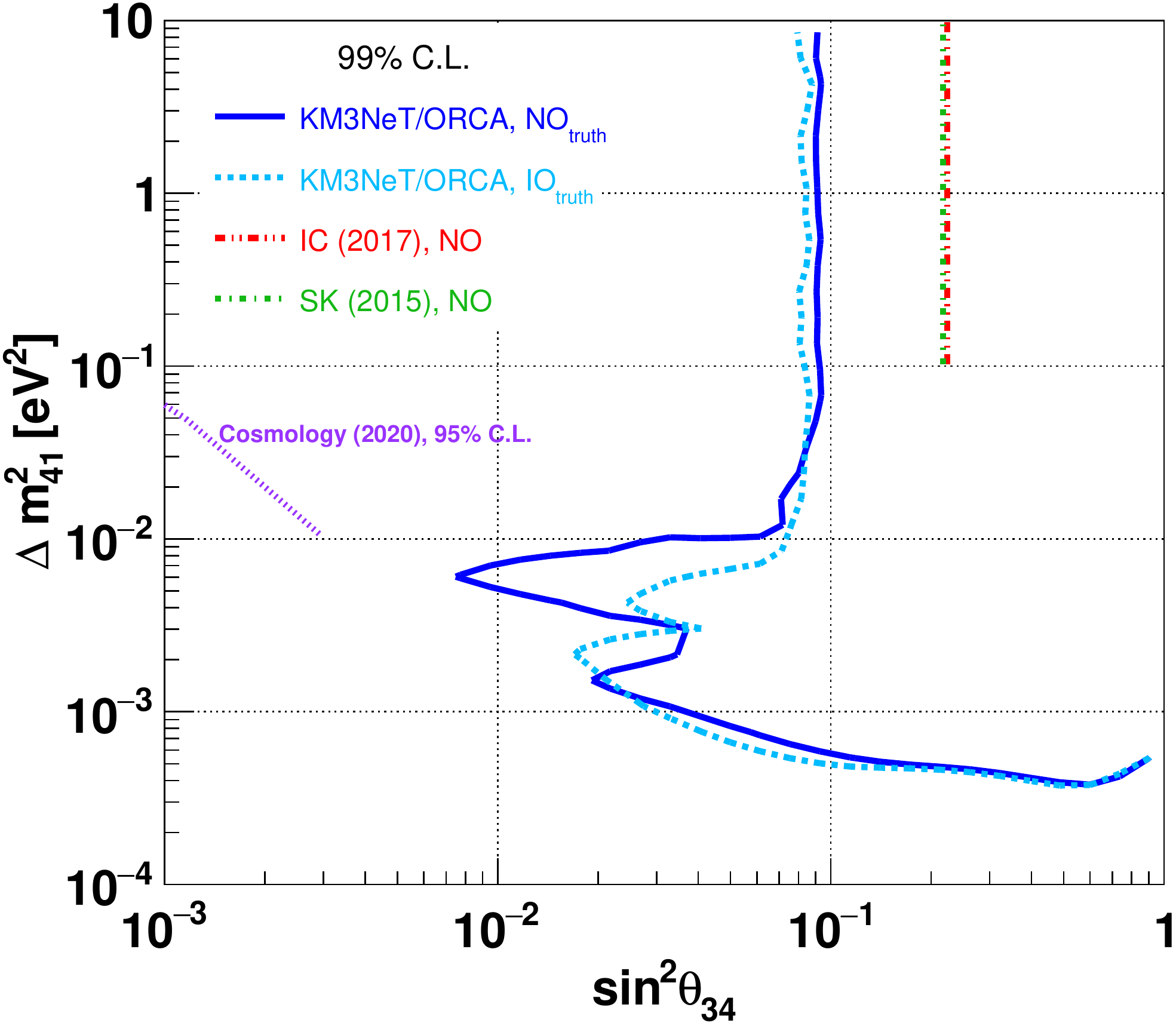}
\mycaption{The $99\%$ C.L. KM3NeT/ORCA sensitivity to the mixing parameter $\theta_{34}$, for different values of $\dmf$, for three years of data taking. Sensitivity results are compared with current upper limits from cosmology \cite{strongboundCosmology}, IceCube/DeepCore \cite{DeepCore_Steriles} and SK \cite{SK}. The excluded region is the one on the right of the lines.}
\label{fig:sen-Ut}
\end{figure}

Fig. \ref{fig:sen-Ut} shows the ORCA sensitivity at $99\%$ C.L. to $\sin^2\theta_{34}$ after three years of data taking. Here, $\theta_{14}$, $\theta_{24}$, $\delta_{14}$ and $\delta_{24}$ are set free in the fit. Upper limits from cosmology \cite{strongboundCosmology}, IceCube/DeepCore \cite{DeepCore_Steriles} and SK \cite{SK} are also reported. In the LF region there are no upper limits on $\theta_{34}$ coming from other experiments. 
\\
ORCA is able to constrain $\theta_{34}$ over a broad range of $\dmf$. In the HF region, consistently with Fig. \ref{fig:sen-Um-Ut}, ORCA can improve current upper limits on $\sin^2\theta_{34}$ by about a factor two. 

\section{Summary and Conclusions}
\label{sec:summary}
KM3NeT/ORCA, a neutrino detector under construction in the Mediterranean Sea, is optimised for oscillation studies with atmospheric neutrinos in the GeV energy range. In this paper, it has been shown that the ORCA detector has a great potential to search for the presence of a light sterile neutrino in the range $\dmf \in [10^{-5}, 10] \, \eV$, by fitting the expected number of observed events classified in three topologies, namely track, intermediate and shower events. With this methodology, ORCA can probe regions in the active-sterile mixing elements $\theta_{14}$, $\theta_{24}$, $\theta_{34}$ and the effective parameter $\theta_{\mu e}$, not yet constrained by current experiments. Particularly, after three years of data taking, ORCA can improve current limits on $\sin^2\theta_{34} \cos^2 \theta_{24}$ by about a factor of two, in case of null result, for an eV-mass sterile neutrino. For lower sterile neutrino masses, down to $\dmf \rightarrow 10^{-5}$ $\eV$, ORCA will be able to test the unexplored region of the $\sin^2\theta_{24}$ parameter, and $\sin^2 2\theta_{\mu e}$ effective parameter space down to about two orders of magnitude with respect to current limits. The ORCA sensitivity to $\sin^2\theta_{14}$ is comparable to current upper limits. Finally, in case of null result, ORCA will able to improve current limits on $\sin^2\theta_{34}$ by about a factor two for an eV-mass sterile neutrino, and it is the first experiment, to date, able to constrain $\theta_{34}$ in the very low sterile mass region. 

%%%%%%%%%%%%%%%%%%%%%%%%%%%%%%%%%%%%%%%%%%%%%%%%%%%%%%%%%%%%%%%%

\section{Acknowledgements} The authors acknowledge the financial support of the funding agencies: Agence Nationale de la Recherche (contract ANR-15-CE31-0020), Centre National de la Recherche Scientifique (CNRS), Commission Europ\'eenne (FEDER fund and Marie Curie Program), Institut Universitaire de France (IUF), LabEx UnivEarthS (ANR-10-LABX-0023 and ANR-18-IDEX-0001), Paris \^Ile-de-France Region, France; Shota Rustaveli National Science Foundation of Georgia (SRNSFG, FR-18-1268), Georgia; Deutsche Forschungsgemeinschaft (DFG), Germany; The General Secretariat of Research and Technology (GSRT), Greece; Istituto Nazionale di Fisica Nucleare (INFN), Ministero dell'Universit\`a e della Ricerca (MIUR), PRIN 2017 program (Grant NAT-NET 2017W4HA7S) Italy; Ministry of Higher Education Scientific Research and Professional Training, ICTP through Grant AF-13, Morocco; Nederlandse organisatie voor Wetenschappelijk Onderzoek (NWO), the Netherlands; The National Science Centre, Poland (2015/18/E/ST2/00758); National Authority for Scientific Research (ANCS), Romania; Ministerio de Ciencia, Innovaci\'{o}n, Investigaci\'{o}n y Universidades (MCIU): Programa Estatal de Generaci\'{o}n de Conocimiento (refs. PGC2018-096663-B-C41, -A-C42, -B-C43, -B-C44) (MCIU/FEDER), Generalitat Valenciana: Prometeo (PROMETEO/2020/019), Grisol\'{i}a (ref. GRISOLIA/2018/119) and GenT (refs. CIDEGENT/2018/034, /2019/043, /2020/049) programs, Junta de Andaluc\'{i}a (ref. A-FQM-053-UGR18), La Caixa Foundation (ref. LCF/BQ/IN17/11620019), EU: MSC program (ref. 101025085), Spain.

\bibliography{biblio}

\end{document}